\documentclass[
    ,final            
  ]
  {aipproc}

\layoutstyle{6x9}

\begin{document}

\title{Benchmark calculation of no-core Monte Carlo shell model in light nuclei}


\author{T.~Abe}{
  address={Department of Physics, the University of Tokyo, Hongo, Tokyo 113-0033, Japan}
}

\author{P.~Maris}{
  address={Department of Physics and Astronomy, Iowa State University, Ames, Iowa 50011, USA}
}

\author{T.~Otsuka$^{\ast,}$}{
  address={CNS, the University of Tokyo, Hongo, Tokyo 113-0033, Japan}
  ,altaddress={NSCL, Michigan State University, East Lansing, Michigan 48824, USA}
}

\author{N.~Shimizu}{
  address={Department of Physics, the University of Tokyo, Hongo, Tokyo 113-0033, Japan}
}

\author{Y.~Utsuno}{
  address={ASRC, Japan Atomic Energy Agency, Tokai, Ibaraki 319-1195, Japan}
}

\author{J.~P.~Vary}{
  address={Department of Physics and Astronomy, Iowa State University, Ames, Iowa 50011, USA}
}

\begin{abstract}
The Monte Carlo shell model is firstly applied to the calculation of the no-core shell model in light nuclei. The results are compared with those of the full configuration interaction. The agreements between them are within a few $\%$ at most. 
\end{abstract}

\maketitle


\section{Introduction}

One of the major challenges of nuclear theory is to understand the nuclear structure from {\it ab-inito} calculations with realistic nuclear forces. The no-core shell model (NCSM) \cite{ref1} is one of these {\it ab-initio} methods. One obstacle for carrying out these calculations is the demand for extensive computational resources. Even at state-of-the-art computational facilities, the NCSM calculations are restricted up to the p-shell nuclei with sufficiently large model spaces \cite{ref2}. Therefore, a method to overcome the current computational limitation of the standard NCSM calculations is needed. 

With this motivation, the Monte Carlo shell model (MCSM) \cite{ref3} is applied to no-core full configuration interaction (FCI) calculations. The MCSM is based on the idea of stochastic sampling of the bases. We can reduce the large Hamiltonian matrices and diagonalize the smaller matrices spanned by a small number of importance-truncated bases stochastically selected. In such a way, we can carry out calculations comparable to the large-scale diagonalization in the standard NCSM calculations. 

\section{Numerical Results}

As the benchmark, we select 9 states of light nuclei; $^4$He ($0^+$), $^6$He ($0^+$), $^6$Li ($1^+$), $^7$Li ($1/2^-$, $3/2^-$), $^8$Be ($0^+$), $^{10}$B ($1^+$, $3^+$), and $^{12}$C ($0^+$). 
The calculated observables are the binding energy, the point-particle root-mean-square (RMS) matter radius, and the electromagnetic moments. 
The model space is truncated by the number of the major shells for the single-particle states. We adopt $N_{shell} = 2$, $3$, and $4$ in this work. 
The oscillator energy, $\hbar \Omega$, is optimized to give the lowest energy for that state and model space. 
The effects of the Coulomb force and the spurious center-of-mass excitation are not considered for this benchmark. 
The MCSM results are compared with those of FCI, which gives the exact results in the chosen single-particle model space. The FCI results are obtained by the MFDn code \cite{ref4}, and the MCSM results by the newly developed code \cite{ref5}. We extrapolate the MCSM results of the energy and the radius by using the energy variance, which is a new ingredient of the recent MCSM approach \cite{ref6}. 
Both in the MCSM and FCI calculations, we use the JISP16 $NN$ interaction \cite{ref7}. 

\begin{figure}[htbp]
    \begin{tabular}{cc}
      \resizebox{75mm}{!}{\includegraphics{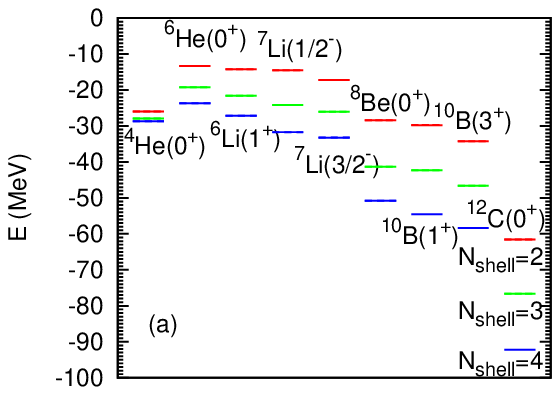}} &
      \resizebox{75mm}{!}{\includegraphics{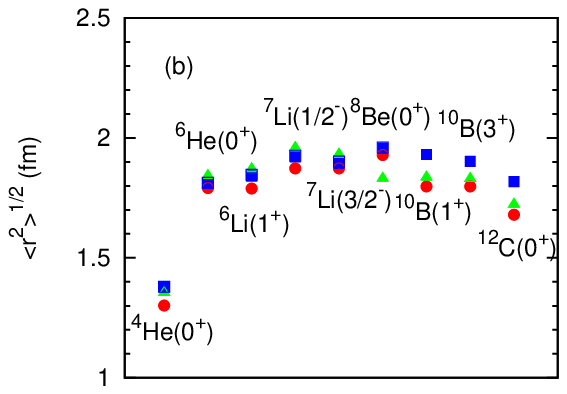}} \\
      \resizebox{75mm}{!}{\includegraphics{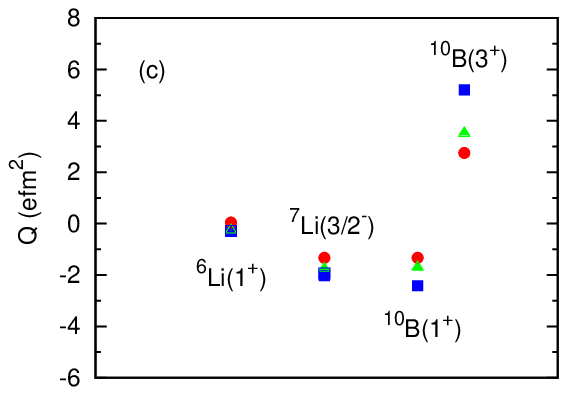}} &
      \resizebox{75mm}{!}{\includegraphics{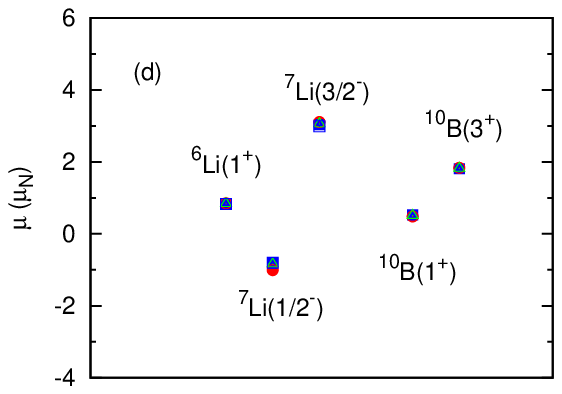}} \\
    \end{tabular}
    \caption{(Color online) Comparisons of (a) the binding energies, (b) the point-particle RMS matter radii, (c) the electric quadrupole and (d) the magnetic dipole moments between the MCSM and FCI. (a) For binding energies, the MCSM (FCI) results are shown as the solid (dashed) lines. From top to bottom, the truncation of the model space is $N_{shell} =$ $2$ (red), $3$ (green), and $4$ (blue). (b), (c), and (d) For these observables, the solid (open) symbols stand for the MCSM (FCI) results. The circle (red), the triangle (green), and the square (blue) indicate the results at $N_{shell} =$ $2$, $3$, and $4$, respectively. Note that all of the results of $^{10}$B and $^{12}$C at $N_{shell} = 4$ were performed only by MCSM.}
    \label{fig1}
\end{figure}

The binding energies obtained by MCSM and FCI can be found in FIGURE 1 (a). The MCSM results are extrapolated by the energy variance with the second-order polynomials. In the figure, the MCSM results are compared with the exact results by FCI. As shown in TABLE 1, the differences between them are around a few tens of keV at most, and cannot be recognized at the energy scale of the figure. The binding energies of $^{10}$B ($1^+$, $3^+$) and $^{12}$C ($0^+$) are the predictions by MCSM. These calculations exceed the current computational limitation of FCI. 
The point-particle RMS matter radii are shown in FIGURE 1 (b). The MCSM results are extrapolated by the energy variance with the first-order polynomials. The differences between the MCSM and FCI are quite small, and are roughly of the order of $10^{-3}$ fm at most. 
FIGURE 1 (c) and (d) show the MCSM and FCI results of the electromagnetic moments. For these observables, due to the large cancellation of the contaminations from the excited states, even without the energy-variance extrapolations, the MCSM gives results within a few $\%$ of the exact FCI results. Note that for the magnetic moments the dependence on the model space is quite small both for MCSM and FCI results. 
TABLE 1 summarizes the MCSM and FCI results of various observables. All of the MCSM results with the energy-variance extrapolations give sufficiently converged results that agree with the FCI results to within a few $\%$. 

\begin{table}[htbp]
\caption{Binding energies, point-particle RMS matter radii, and electromagnetic moments.} \label{table1}
    \begin{tabular}{cccccc}
      \hline
      Nuclei\tablenote{The quantum numbers inside the parentheses after the atomic symbols are the spin and parity.} & Method & E (MeV) \tablenote{For the entries of MCSM, the quantities inside the parentheses are $N_{shell}$,  $\hbar \Omega$ (MeV), and the number of MCSM dimensions, while for FCI, $N_{shell}$ and $\hbar \Omega$ (MeV).} \ $^{, \ast \ast}$ & $\sqrt{\langle r^2 \rangle}$ (fm) \tablenote{The bars in the entries for FCI results denote the results are not available, so far.} \ \ $^{, \ddagger}$ & Q ($e$fm$^2$)$^{\ast \ast ,}$ \tablenote{The entries in the column are the results at $N_{shell} = 2$, $3$, and $4$, respectively.} \ $^{, \S}$ & $\mu$ ($\mu_N$)$^{\ast \ast , \ddagger ,}$ \tablenote{The entries left blank indicate the values are exactly $0$.}\\
      \hline
        $^4$He ($0^+$)   &MCSM & -25.956 (2,30,4), -27.914 (3,30,38), \ \ \ -28.738 (4,30,60) & 1.301, 1.355, 1.379 & & \\
                         &FCI & -25.956 (2,30), \ \ \ -27.914 (3,30), \ \ \ \ \ \ \ \ -28.738 (4,30) \ \ \ \ \ & 1.301, 1.355, 1.379 & & \\
        $^6$He ($0^+$)   &MCSM& -13.343 (2,20,7), -19.196 (3,20,77), \ \ \ -23.701 (4,25,67) & 1.791, 1.843, 1.813 & & \\
                         &FCI & -13.343 (2,20), \ \ \ -19.196 (3,20), \ \ \ \ \ \ \ \ -23.684 (3,25) \ \ \ \ \ & 1.791, 1.843, 1.813 & & \\
        $^6$Li ($1^+$)   &MCSM& -14.218 (2,20,8), -21.581 (3,20,89), \ \ \ -27.168 (4,25,66) & 1.789, 1.871, 1.846 & 0.044, -0.260, -0.280 & 0.852, 0.836, 0.835 \\
                         &FCI & -14.218 (2,20), \ \ \ -21.581 (3,20), \ \ \ \ \ \ \ \ -27.168 (4,25) \ \ \ \ \ & 1.789, 1.871, 1.846 & 0.043, -0.259, -0.285 & 0.852, 0.835, 0.832 \\
        $^7$Li ($1/2^-$) &MCSM& -14.459 (2,20,10), -24.167 (3,20,103), -31.705 (4,25,70) & 1.873, 1.959, 1.926 & & -1.009, -0842, -0.815 \\
                         &FCI & -14.458 (2,20), \ \ \ \ \ -24.165 (3,20), \ \ \ \ \ \ \ -31.748 (4,25) \ \ \ \ \ & 1.873, 1.959, 1.926 & & -1.009, -0.840, -0.807 \\
        $^7$Li ($3/2^-$) &MCSM& -17.232 (2,20,10), -26.064 (3,25,100), -33.276 (4,25,65) & 1.874, 1.932, 1.899 & -1.328, -1.772, -2.025 & 3.109, 3.064, 3.036 \\
                         &FCI & -17.232 (2,20), \ \ \ \ \  -26.063 (3,25), \ \ \ \ \ \ \ -33.202 (4,25) \ \ \ \ \ & 1.873, 1.932, 1.901 & -1.328, -1.750, -1.940 & 3.109, 3.056, 2.993 \\
        $^8$Be ($0^+$)   &MCSM& -28.435 (2,20,7), -41.291 (3,25,57), \ \ \  -50.756 (4,25,58) & 1.929, 1.831, 1.957 & & \\
                         &FCI & -28.435 (2,20), \ \ \ -41.291 (3,25), \ \ \ \ \ \ \ \ -50.756 (4,25) \ \ \ \ \ & 1.929, 1.831, 1.960 & & \\
      $^{10}$B ($1^+$)   &MCSM& -29.755 (2,25,9), -42.331 (3,25,92), \ \ \ -54.812 (4,25,76) & 1.798, 1.837, 1.958 & -1.333, -1.715, -2.416 & 0.483, 0.503, 0.526 \\
                         &FCI & -29.755 (2,25), \ \ \ -42.338 (3.25), \ \ \ \ \ \ \ \ --------------------- & 1.798, 1.836, ------- & -1.333, -1.698, ------- & 0.486, 0.509, ------- \\
      $^{10}$B ($3^+$)   &MCSM& -34.221 (2,25,5), -46.602 (3.25,77), \ \ \ -58.345 (4,25,56)& 1.798, 1.831, 1.924 & 2.750, 3.554, 5.204 & \ 1.836, 1.820, \ 1.813 \ \\
                         &FCI & -34.221 (2,25), \ \ \ -46.602 (3,25), \ \ \ \ \ \ \ \ --------------------- & 1.798, 1.830, ------- & \ 2.750, \ 3.503, \ ------ & 1.836, 1.818, ------- \\
      $^{12}$C ($0^+$)   &MCSM& -62.329 (2,30,4), -76.621 (3,30,78), \ \ \ -92.179 (4,30,82)& 1.680, 1.723, 1.818 & & \\
                         &FCI & -62.329 (2,30), \ \ \ -76.621 (3,30), \ \ \ \ \  \ \ \ --------------------- & 1.680, 1.723, ------- & & \\
      \hline
    \end{tabular}
\end{table}

\section{Summary}

As the exploratory work, the MCSM approach has been applied to the FCI calculations. We have performed the benchmark of the binding energies, point-particle RMS matter radii, and the electromagnetic moments for light nuclei ranging from $^4$He to $^{12}$C. The binding energies and the point-particle RMS matter radii calculated by MCSM were extrapolated by the energy variance. The MCSM and FCI results were compared in various model spaces. The no-core MCSM shows good agreement with the FCI results within a few $\%$ of deviation at most for the observables we have investigated. The details of this work can be found in future publications \cite{ref8}.


\begin{theacknowledgments}
This work was supported by Grants-in-Aid for Young Scientists (No.~20740127 and No.~21740204), for Scientific Research (No.~20244022), and for Scientific Research on Innovative Areas (No.~20105003) from JSPS, and by the CNS-RIKEN joint project for large-scale nuclear structure calculations. It was also supported in part by the US DOE Grants DE-FC02-07ER41457, DE-FC02-09ER41582 (UNEDF SciDAC Collaboration) and DE-FG02-87ER40371, and through JUSTIPEN under grant number DEFG02-06ER41407. A part of the MCSM calculations was performed on the T2K Open Supercomputer at the University of Tokyo and University of Tsukuba, and the BX900 Supercomputer at JAEA. A part of the FCI calculations was performed on the supercomputers at NERSC and at ORNL (INCITE Award). 
\end{theacknowledgments}

\end{document}